\documentclass[ste,twoside]{stefano}

\usepackage{amsmath}
\usepackage{amssymb}
\usepackage{graphics}
\usepackage{rotating}
\usepackage{cite}
\usepackage{color}

\def\makeheadbox{{%
\hbox to0pt{\vbox{\baselineskip=10dd\hrule\hbox
to\hsize{\vrule\kern3pt\vbox{\kern3pt \hbox{{\sf Journal of
Physics A} {\bf 38}, 3443-3454 (2005) }
\hbox{
{\sc {\color{blue}{dma}}[{\color{black}{imecc}}]{\color{red}{UniCamp}}
}
\hspace*{10.3cm}
{\color{blue}{$\boldsymbol{\Sigma \delta \Lambda}$ }}}
\kern3pt}\hfil\kern3pt\vrule}\hrule}%
\hss}}}

\def\0{\mbox{\tiny $0$}}
\def\1{\mbox{\tiny $1$}}
\def\2{\mbox{\tiny $2$}}
\def\3{\mbox{\tiny $3$}}
\def\4{\mbox{\tiny $4$}}
\def\5{\mbox{\tiny $5$}}
\def\6{\mbox{\tiny $6$}}
\def\7{\mbox{\tiny $7$}}
\def\8{\mbox{\tiny $8$}}
\def\9{\mbox{\tiny $9$}}
\def\m{\mbox{\tiny $-$}}
\def\p{\mbox{\tiny $+$}}
\def\mip{\mbox{\tiny $\pm$}}
\def\ov{\mbox{\tiny $/$}}
\def\R{\mbox{\tiny $\mathbb{R}$}}
\def\C{\mbox{\tiny $\mathbb{C}$}}
\def\H{\mbox{\tiny $\mathbb{H}$}}
\def\I{\mbox{\tiny I}}
\def\II{\mbox{\tiny II}}

\begin{document}
%

\title{QUATERNIONIC BOUND STATES}

\author{
Stefano De Leo\inst{1}
\and
Gisele C. Ducati\inst{2}
}

\institute{
Department of Applied Mathematics, University of Campinas\\
PO Box 6065, SP 13083-970, Campinas, Brazil\\
{\em deleo@ime.unicamp.br}
 \and
Department of Mathematics, University of Parana\\
PO Box 19081, PR 81531-970, Curitiba, Brazil\\
{\em ducati@mat.ufpr.br}
}


\date{Submitted: {\em October, 2004}.  Revised: {\em February, 2005}.}

\abstract{We study the bound-state solutions of vanishing angular
momentum in a quaternionic spherical square-well potential of
finite depth. As in the standard quantum mechanics, such solutions
occur for discrete values of energies. At first glance, it seems
that the continuity conditions impose a very restrictive
constraint on the energy eigenvalues and, consequently, no bound
states was expected for energy values below the pure quaternionic
potential. Nevertheless, a careful analysis shows that pure
quaternionic potentials do not remove bound states. It is also
interesting to compare these new  solutions with the bound state
solutions of the trial-complex potential. The study presented in
this paper represents a preliminary step towards a full
understanding of the role that quaternionic potentials could play
in quantum mechanics. Of particular interest for the authors it is
the analysis of confined wave packets and tunnelling times in this
new formulation of quantum theory.}


\PACS{ {02.30.Tb} \and {03.65.Ca}{}}










\titlerunning{Quaternionic bound states}

\maketitle


\section*{I. INTRODUCTION}

In this introductory section, we briefly recall the basic
kinematic and dynamical framework of quaternionic quantum
mechanics. Obviously, it is not our purpose to give a complete and
satisfactory introduction to the subject. For this and for further
details on the mathematical formalism used in approaching a
systematic development of quaternionic quantum mechanics, covering
an axiomatic mathematical introduction to quaternionic quantum
field and an interesting  discussion on the role that quaternions
could play in searching for new physics, we refer the reader to
the excellent book of Adler\cite{ADL}.

In last years, the study of quaternionic quantum mechanics shed
new light on the mathematical structures used in quantum
mechanics. In particular, a better understanding of the eigenvalue
problem for real/complex and quaternionic linear differential
operators\cite{EP1,EP2} allowed the solution of the Schr\"odinger
equation in presence of quaternionic potentials\cite{QDO} without
(unnecessary) complex translations\cite{DAV89,DAV92}. Nonetheless,
quaternionic quantum mechanics is still in the early stage of
development and what is still lacking is an interpretation of
quaternionic potentials and proposals of experimental tests for
quaternionic deviations of standard quantum mechanics. According
to the nonrelativistic quaternionic scattering theory (developed
in detail by Adler\cite{ADL}) and the recent work on quaternionic
tunneling effect\cite{QTE}, experiments to detect left/right
transmission asymmetry in presence of quaternionic barriers should
involve CP-violating (asymmetric) potentials. Consequently, the
earliest proposals for tests of quaternionic quantum mechanics
based on neutron-optical experiments gave (as expected in virtue
of the study on quaternionic scattering theory) a null
result\cite{PER,KAI,KLE}. It is thus important to look for other
possible signatures of violation of the standard quantum mechanics
due to quaternionic potentials. The main scope of this paper is to
solve the quaternionic Schr\"odinger equation in presence of
binding potentials. This could represent a potential (and not yet
explored) research field to determine if and where quaternionic
potentials could be seen.

\section*{II. MATHEMATICAL TOOLS IN QUATERNIONIC QUANTUM MECHANICS}

To make our exposition self-contained and to facilitate access to
the individual topics, the sections are rendered self-contained
and, when necessary, we repeat the relevant material recently
appeared in litterature. Before to begin the study of quaternionic
binding potentials, we summarize our notation and introduce some
basic concepts of quaternionic quantum mechanics.\\

\noindent
{\bf $\bullet$ Right multiplication by quaternionic scalars.}\\
 In close analogy with the familiar complex quantum mechanics,
the states of quaternionic quantum mechanics will be described by
wave functions belong to an abstract vector space. Due to the
noncommutative nature of the quaternionic multiplication, we have
to specify whether the quaternionic Hilbert space,
$\mathcal{V}_{\mathbb{H}}$, is linear under right or left
multiplication by quaternionic scalars. Following the Adler
book\cite{ADL}, we adopt the convention of linearity from the
right for the quaternionic Hilbert space. Thus, if
$\Phi_{\H,\1}(\boldsymbol{r,t})$ and
$\Phi_{\H,\2}(\boldsymbol{r,t})$ belong to
$\mathcal{V}_{\mathbb{H}}$ and $q_{\1}$ and $q_{\2}$ are
quaternionic scalars, then
\begin{equation}
\Phi_{\H,\1}(\boldsymbol{r},t) \, q_{\1} +
\Phi_{\H,\2}(\boldsymbol{r},t) \, q_{\2}
~\in~~\mathcal{V}_{\mathbb{H}}~.
\end{equation}

\noindent
{\bf $\bullet$ Anti-self adjoint operators and Schr\"odinger equation.}\\
The time evolution of $\Phi_{\H}(\boldsymbol{r},t)$ is governed by
the {\em quaternionic} Scr\"odinger equation:
\begin{equation}
\label{qse}
\partial_t \, \Phi_{\H}(\boldsymbol{r},t) =
\mathcal{A}_{\H}(\boldsymbol{r},t)\, \Phi_{\H}(\boldsymbol{r},t)~,
\end{equation}
where
\begin{eqnarray*}
\mathcal{A}_{\H}(\boldsymbol{r},t) & = & \left[ \, i \,
\mbox{$\frac{\hbar^{\2}}{2m}$} \, \nabla^{\2}  - i \,
V_{\R,\1}(\boldsymbol{r},t) - j \, V_{\R,\2}(\boldsymbol{r},t) - k
\,
V_{\R,\3}(\boldsymbol{r},t) \, \right] / \,\hbar\\
& = & \mathcal{A}_{\C}(\boldsymbol{r},t) + j \,
\mathcal{B}_{\C}(\boldsymbol{r},t)~,
\end{eqnarray*}
is the anti-self-adjoint operator associated with the total energy
of the system. The introduction of a pure quaternionic potential,
$j \, \mathcal{B}_{\C}(\boldsymbol{r},t)$, does not modify the
transition probability. There is an important difference between
complex and quaternionic quantum mechanics. In complex quantum
mechanics, we can trivially relate any anti-self-adjoint operator
$\mathcal{A}_{\C}(\boldsymbol{r},t)$ to the Hamiltonian observable
$\mathcal{H}_{\C}(\boldsymbol{r},t)$ by the  factor $i$, namely
\[
\mathcal{H}_{\C}(\boldsymbol{r},t) = i \, \hbar \,
\mathcal{A}_{\C}(\boldsymbol{r},t)~.\] This is obviously not true
in quaternionic quantum mechanics. In fact, the left-action of the
complex imaginary unit $i$ on  $j \,
\mathcal{B}_{\C}(\boldsymbol{r},t)$ {\em still} gives a pure
quaternionic potential, $k \,
\mathcal{A}_{\C}(\boldsymbol{r},t)$.\\

\noindent
{\bf $\bullet$ Stationary state and eigenvalue problem}\\
In the presence of time-independent potentials, the quaternionic
Schr\"odinger equation (\ref{qse}) becomes
\begin{eqnarray}
\label{tiqse}
\partial_t \, \Phi_{\H}(\boldsymbol{r},t) & = &
\mathcal{A}_{\H}(\boldsymbol{r})\, \Phi_{\H}(\boldsymbol{r},t)
\nonumber \\
 & = &
\left[ \, i \, \mbox{$\frac{\hbar^{\2}}{2m}$} \, \nabla^{\2} - i
\, V_{\R,\1}(\boldsymbol{r}) - j \, V_{\R,\2}(\boldsymbol{r}) - k
\, V_{\R,\3}(\boldsymbol{r}) \, \right] \,
\Phi_{\H}(\boldsymbol{r},t) \,/ \, \hbar
\end{eqnarray}
The method of separation of variables is still applicable if we
set
\[\Phi_{\H}(\boldsymbol{r},t)= \varphi_{\H}(\boldsymbol{r})\,
\chi_{\H}(t)~.
\]
Due to the linearity under right multiplication by quaternionic
scalars of $\mathcal{A}_{\H}(\boldsymbol{r})$, the right position
of the time-dependent quaternionic function $\chi_{\H}(t)$ plays a
fundamental role in separating the initial differential equation
in two equations containing respectively the time and space
variables. Explicitly, we find
\[
\hbar\,\dot{\chi}_{\H}(t) \, \chi^{\m \1}_{\H}(t) = \varphi^{\m
\1}_{\H}(\boldsymbol{r}) \, \left[ \, i \,
\mbox{$\frac{\hbar^{\2}}{2m}$} \, \nabla^{\2} - i\,
V_{\R,\1}(\boldsymbol{r}) - j \, V_{\R,\2}(\boldsymbol{r}) - k \,
V_{\R,\3}(\boldsymbol{r}) \, \right] \,
\varphi_{\H}(\boldsymbol{r})\, \, .
\]
This equality is only possible if each of these functions is a
constant, which we shall set equal to $\lambda \in \mathbb{H}$.
Thus, the right linearity of the quaternionic Hilbert space
$\mathcal{V}_{\H}$ implies a right eigenvalue problem for the
anti-self adjoint operator $\mathcal{A}_{\H}(\boldsymbol{r})$,
\begin{equation}
\label{eql} \mathcal{A}_{\H}(\boldsymbol{r})\,
\varphi_{\H}(\boldsymbol{r}) = \varphi_{\H}(\boldsymbol{r}) \,
\lambda \, /\,\hbar~.
\end{equation}
The need to use right eigenvalue in physical problems can also be
demonstrate by simple mathematical arguments\cite{EP1}.\\

\noindent
{\bf $\bullet$ Number systems and amplitudes of probabilities.}\\
To give a satisfactory probability interpretation, amplitudes of
probability must be defined in associative division algebras.
Amplitudes of probabilities defined in non-division algebras fails
to satisfy the requirement that in the absence of quantum
interference effects, probability amplitude superposition should
reduce to probability superposition. The associative law of
multiplication (which fails for the octonions) is needed to
satisfy the completeness formula and to guarantee that the
Sch\"odinger anti-self-adjoint operator leaves invariant the inner
product. The important point to note here is that we can still use
vectors in Clifford algebraic or octonionic Hilbert space.
Nevertheless, amplitudes of probability must be defined in
$\mathbb{C}$ or $\mathbb{H}$. For example, we can formulate a
consistent complexified quaternionic\cite{CQQM} or
octonionic\cite{OQM} quantum mechanics by adopting complex inner
products\cite{CSP}, i.e. complex projections of complexified
quaternionic or octonionic inner product. The use of complex inner
product represents a fundamental tool in applying a Clifford
algebra formalism to physics and plays a fundamental role in
looking for geometric interpretation of the algebraic structures
in relativistic equations\cite{RE1,RE2,RE3,RE4} and gauge
theories\cite{GT1,GT2}. The choice of quaternionic inner product
seems to be the best adapted to investigate deviations from the
standard complex theory. In this paper, we shall work with {\em
quaternionic} inner products. Thus, with each pair of elements of
the quaternionic Hilbert space $\mathcal{V}_{\mathbb{H}}$, we
associate a quaternionic inner product defined by
\begin{equation}
\langle \, \varphi_{\H} \, , \, \psi_{\H} \, \rangle ~:=~ \int
\mbox{d} \boldsymbol{r} \, \,  \,
\overline{\varphi}_{\H}(\boldsymbol{r}) \,
\,\psi_{\H}(\boldsymbol{r})~~~~\in \mathbb{H},
\end{equation}
where $\overline{\varphi}_{\H}(\boldsymbol{r})$ denotes the
quaternionic conjugate of
\[
\begin{array}{lcl} \varphi_{\H}(\boldsymbol{r}) & = &
\varphi_{\R,\0}(\boldsymbol{r}) + i \,
\varphi_{\R,\1}(\boldsymbol{r}) + j \,
\varphi_{\R,\2}(\boldsymbol{r}) + k \,
\varphi_{\R,\3}(\boldsymbol{r})~,
\end{array}
\]
i.e.
\[
\begin{array}{lcl} \overline{\varphi}_{\H}(\boldsymbol{r}) & = &
\varphi_{\R,\0}(\boldsymbol{r}) - i \,
\varphi_{\R,\1}(\boldsymbol{r}) - j \,
\varphi_{\R,\2}(\boldsymbol{r}) - k \,
\varphi_{\R,\3}(\boldsymbol{r})~.
\end{array}
\]
\\

\noindent
{\bf $\bullet$ Energy eigenvalues.}\\
The anti-hermiticity of $\mathcal{A}_{\H}(\boldsymbol{r})$,
\[ \langle \, \varphi_{\H} \, , \,  \mathcal{A}_{\H} \, \psi_{\H} \, \rangle =
- \, \langle \, \mathcal{A}_{\H} \,\varphi_{\H} \, , \, \psi_{\H}
\, \rangle
\] implies $\lambda = - \overline{\lambda}$. Consequently, observing that
$\lambda$ is related to the total energy of the system and setting
\[ \lambda = - \left(iE_{\1} + jE_{\2}+kE_{\3}\right)~, \]
the quaternionic energy eigenvalue equation (\ref{eql}) becomes
\begin{equation}
\label{eqq} \mathcal{A}_{\H}(\boldsymbol{r})\,
\varphi_{\H}(\boldsymbol{r}) =- \, \varphi_{\H}(\boldsymbol{r}) \,
(iE_{\1}+jE_{\2}+kE_{\3})\,/\,\hbar~.
\end{equation}
In the complex limit, the previous equation reduces to
\[
 \mathcal{H}_{\C}(\boldsymbol{r}) \, \varphi_{\C}(\boldsymbol{r}) =
i \, \hbar \, \mathcal{A}_{\C}(\boldsymbol{r},t)\,
\varphi_{\C}(\boldsymbol{r}) = E_{\1} \,
\varphi_{\C}(\boldsymbol{r}) \] Eigenvectors of anti-self-adjoint
quaternionic operator divide into mutually orthogonal
eigenclasses. Each eigenclass corresponds to a ray of physically
equivalent states. Eigenvectors within each eigenclass are not
orthogonal and have eigenvalues related by the automorphism
transformation
\[
\bar{u} \, \left(i \, \lambda_{\1} + j \, \lambda_{\2} + k \,
\lambda_{\3} \right) \, u~,
\]
with $u$ unitary quaternion. By choosing a particular automorphism
transformation, i.e.
\[
\omega = \mbox{$\sqrt{ \frac{\lambda_{\1} +
\sqrt{\lambda_{\1}^{^{\2}} + \lambda_{\2}^{^{\2}}
+\lambda_{\3}^{^{\2}}}}{ 2 \, \sqrt{\lambda_{\1}^{^{\2}} +
\lambda_{\2}^{^{\2}} +\lambda_{\3}^{^{\2}}}}}$}
 \, \left[ \, 1 + j \, \frac{\lambda_{\2} - i \lambda_{\3}}{i \,
\left( \lambda_{\1} + \sqrt{\lambda_{\1}^{^{\2}} +
\lambda_{\2}^{^{\2}} +\lambda_{\3}^{^{\2}}} \, \right)} \,
\right]~,
\]
we set a particular representative ray within each eigenclass
whose eigenvalue is complex,
\[ \lambda = i \,
\sqrt{\lambda_{\1}^{^{\2}}+
\lambda_{\2}^{^{\2}}+\lambda_{\3}^{^{\2}}}~.
\]
Hence, within each eigenclass of
$\mathcal{A}_{\H}(\boldsymbol{r})$ eigenvectors, there is a ray
\[
\Psi_{\H}(\boldsymbol{r}) = \varphi_{\H}(\boldsymbol{r}) \, u
\]
for which
\begin{equation}
 \mathcal{A}_{\H}(\boldsymbol{r})\,
\Psi_{\H}(\boldsymbol{r}) = - \, \Psi_{\H}(\boldsymbol{r}) \, i \,
E\,/\,\hbar~.
\end{equation}
Thus, for stationary states, the quaternionic Schr\"odinger
equation can be rewritten as follows
\begin{equation}
\label{eqc}
 \left[
\, i \, \mbox{$\frac{\hbar^{\2}}{2m}$} \, \nabla^{\2} - i \,
V_{\R,\1}(\boldsymbol{r}) - j \, V_{\R,\2}(\boldsymbol{r}) - k \,
V_{\R,\3}(\boldsymbol{r}) \, \right] \, \Psi_{\H}(\boldsymbol{r})
= -  \,\Psi_{\H}(\boldsymbol{r}) \, i \, E~.
\end{equation}

\section*{III. SPHERICALLY SYMMETRIC POTENTIALS}

The central force problem in which $V_{\R,\1}(\boldsymbol{r}) =
V_{\R,\1}(r)$ occupies a prominent place in (complex) quantum
mechanics because many important applications involve potentials
that are at least approximately spherically symmetric. In this
section, we extend this symmetry to the quaternionic part of the
potential,
\[
V_{\R,\2}(\boldsymbol{r}) =
V_{\R,\2}(r)~~~\mbox{and}~~~V_{\R,\3}(\boldsymbol{r}) =
V_{\R,\3}(r)~.
\]
Since the quaternionic potential
\[V_{\H}(r)=i \,
V_{\R,\1}(r) + j \, V_{\R,\2}(r) + k \, V_{\R,\3}(r) \,\,,\]
depends only on the distance $r$ of the particle from the origin,
spherical coordinates are the best adapted to the problem. Thus,
we express the Laplace operator in
$\mathcal{A}_{\H}(\boldsymbol{r})$ in spherical coordinates
$(r,\theta,\varphi)$ by the well-known formula
\[
\nabla^{\2} = \frac{\partial^{\, \2}}{\partial r^{\2}} +
\frac{2}{r} \, \frac{\partial}{\partial r}  -
\frac{\boldsymbol{L}^{\2}}{\hbar^{\2} r^{\2}} \, \,  .
\]
Since the operator $\boldsymbol{L}^{\2}$ depends only on $\theta$
and $\varphi$, by using the (complex) spherical harmonic functions
$Y_{\C}^{lm}(\theta,\varphi)$,
\[ \Psi_{\H}(\boldsymbol{r}) = R_{\H}(r) \,
Y_{\C}^{lm}(\theta,\varphi)\,\, ,
\]
we can reduce the Schr\"odinger equation (\ref{eqc}) to the
following {\em quaternionic} radial equation
\begin{equation}
\label{radialR} \left\{ \, i \, \frac{\hbar^{\2}}{2m} \, \left[ \,
\frac{\mbox{d}^{\2}}{\mbox{d}r^{\2}} + \frac{2}{r} \,
\frac{\mbox{d}}{\mbox{d}r} \, \right] -  i \,
V_{\R,\1}^{\mbox{\tiny eff}}(r) - j \, V_{\R,\2}(r) - k \,
V_{\R,\3}(r) \, \right\} \, R_{\H}(r) = - \, R_{\H}(r) \, i \, E~,
\end{equation}
where
\[ V_{\R,\1}^{\mbox{\tiny eff}}(r) =
V_{\R,\1}(r) +   \frac{l(l+1)\,\hbar^{\2} }{2m\,r^{\2}}~.\] By the
change in the function
\[ R_{\H}(r) = U_{\H}(r)\,/\,r~,\]
we obtain for $U_{\H}(r)$ the following differential equation
\begin{equation}
\label{radialUa} \left[ \, i \, \frac{\hbar^{\2}}{2m} \,
\frac{\mbox{d}^{\2}}{\mbox{d}r^{\2}}  - i \,
V_{\R,\1}^{\mbox{\tiny eff}}(r) - j \, V_{\R,\2}(r) - k \,
V_{\R,\3}(r) \, \right] \, U_{\H}(r) = -  \, U_{\H}(r) \, i \, E~.
\end{equation}

\subsection*{III-A. SPHERICAL POTENTIAL TRAP AND QUATERNIONIC EIGENFUNCTIONS}
To find the energy levels of a particle with angular momentum
$l=0$ in presence of a quaternionic spherical potential trap
\[
\begin{array}{lcl}
V_{\H}(r) & = & \left\{ \begin{array}{lll}
0 &~~~~~r<a &\hspace*{1cm} \mbox{\sc (region I)}\, \, ,\\
 & \\
 \,  i\,V_{\1}+j\, V_{\2}+ k \, V_{\3} & ~~~~~r>a &
 \hspace*{1cm}\mbox{\sc (region II)}\, \,
,\end{array} \right.
\end{array}
\]
we have substitute such a potential in Eq.(\ref{radialUa}). This
yields the two ordinary differential equations
\begin{equation}
\label{ereg1} i \, \, \ddot{U}_{\I,\H}(r) = - \, U_{\I,\H}(r) \, i
\, 2mE/\hbar^{\2}~, \end{equation}
and
\begin{equation}
\label{ereg2}
 i \,\ddot{U}_{\I,\H}(r)- [\,2m\,( i \, V_{\1} + j \, V_{\2} + k \, V_{\3})/\hbar^{\2}\,]\,
U_{\II,\H}(r) = - \, U_{\II,\H}(r) \, i \, 2mE/\hbar^{\2}~,
\end{equation}
The solution in region is easily obtained by observing that to the
standard complex sine/cosine solution we have to add a {\em pure}
quaternionic hyperbolic sine/cosine solution,
\[
 U_{\I,\H}(r)  =  \sin (\epsilon \, r) ~\alpha_{\1} + \cos
(\epsilon \, r ) ~ \beta_{\1} + j \, \sinh (\epsilon \, r)
~\gamma_{\1} + j \, \cosh (\epsilon \, r ) ~ \delta_{\1}~,
\]
where $\epsilon = \sqrt{2mE/\hbar^{\2}}$ and
$\alpha_{\1},\,\beta_{\1},\, \gamma_{\1},\, \delta_{\1}$ are
constant complex coefficients to be determined by boundary
conditions and continuity constraints. To guarantee the correct
behavior of the solution at the origin, the wave-function must
satisfy $U_{\I, \H}(0)= 0$, thus we have to impose the following
constraint $\beta_{\1}=\delta_{\1}=0$. Consequently, the solution
inside the well ($r<a$) is given by
\begin{equation}
\label{SII} U_{\I,\H}(r)   =  \sin (\epsilon \, r) ~\alpha_{\1} +
j \, \sinh (\epsilon \, r) ~\gamma_{\1}~.
\end{equation}
The discussion in region II is a little more complicated (see the
appendix A for the explicit calculations) and provides the
following solution
\begin{eqnarray*}
 U_{\II,\H}(r)  & = &   (1+j \, w) \, \left\{ \, \sinh [\nu_{\m}\, r] ~
 \alpha_{\2} + \cosh [\nu_{\m} \, r] ~ \beta_{\2} \, \right\} +
(z + j ) \, \left\{ \, \sinh [\nu_{\p}\, r] ~\gamma_{\2} + \cosh
[\nu_{\p} \, r] ~ \delta_{\2} \, \right\}~,
\end{eqnarray*}
where
\[
\mbox{$\nu_{\mip} = \sqrt{\frac{2m}{\hbar^{\2}} \, \left( V_{\1}
\pm \sqrt{E^{^{\2}}-   V_{\2}^{^{\2}} - V_{\3}^{^{\2}} } \,
\right)}\, \, ,\, \, \,\,\,\,$} \mbox{$w = - \, i \, \frac{V_{\2}
\, - \, i  \, V_{\3}}{E \, + \, \sqrt{E^{^{\2}}-V_{\2}^{^{\2}} -
V_{\3}^{^{\2}} }}\hspace*{.5cm} \mbox{and} \hspace*{.5cm} z = \, i
\, \frac{V_{\2} \, + \, i \, V_{\3}}{E \, + \, \sqrt{E^{^{\2}} -
V_{\2}^{^{\2}} -V_{\3}^{^{\2}} }}$}\, \, .
\]
Observing that, independently of the value of $E$, the real part
of $\nu_{\p }$ is never null, the condition that the wave function
does not diverge at $r=+\infty$  requires a first constraint i.e.
$\gamma_{\2}=-\delta_{\2}$ (without loss of generality we can
choose $\mbox{Re}[\nu_{\mip }]
> 0$). Thus, the solution in region II can be rewritten as
\[
U_{\II,\H}(r)  =  (1+j \, w) \, \left[ \, \sinh (\nu_{\m \,} r)
~\alpha_{\2} + \cosh (\nu_{\m \, } r ) ~ \beta_{\2} \, \right]  +
(z + j ) \, \exp[ - \nu_{\p} \, r] ~\delta_{\2}~,
\]
If the total energy $E$ is greater  than
$\sqrt{V_{\1}^{^{\2}}+V_{\2}^{^{\2}}+V_{\3}^{^{\2}}}$, we have the
free particle case. In fact, for
$E>\sqrt{V_{\1}^{^{\2}}+V_{\2}^{^{\2}}+V_{\3}^{^{\2}}}$,
$\nu_{\m}$ is a complex imaginary number and no constraint exists
for the coefficient $\alpha_{\2}$ and $\beta_{\2}$. The continuity
relations in $r=a$ for the wave-functions and their gradients can
be satisfied for all values of the energy and the energy levels
are therefore not quantized.

The bound particle case corresponds to
$E<\sqrt{V_{\1}^{^{\2}}+V_{\2}^{^{\2}}+V_{\3}^{^{\2}}}$. In this
case, the condition that the wave function does not diverge at
$r=+\infty$ gives an additional constraint on the coefficients
$\alpha_{\2}$ and $\beta_{\2}$, i.e. $\alpha_{\2}=-\beta_{\2}$.
Thus, we have a spatially damped solution in region II given by
\begin{equation}
\label{SIII} U_{\II,\H}\left(r;\mbox{\small
$E<\sqrt{V_{\1}^{^{\2}} + V_{\2}^{^{\2}} + V_{\3}^{^{\2}}}$}
 \, \right)  = (1+j \, w) \, \exp[ - \nu_{\m \, } r] ~\beta_{\2} + (z + j ) \,
\exp[ - \nu_{\p} \, r] ~\delta_{\2}~.
\end{equation}
It remains to adjust the coefficients $\alpha_{\1}$,
$\gamma_{\1}$, $\beta_{\2}$
 and $\delta_{\2}$  so that the amplitudes and
gradients at $r=a$ are continuous. This will imply a quantization
for the energy levels.

\subsection*{III-B. QUATERNIONIC BOUND STATES}
The continuity conditions
\[ U_{\I,\H}(a) = U_{\II,\H}(a)~~~\mbox{and}~~~
 \dot{U}_{\I,\H}(a)=\dot{U}_{\II,\H}(a)
\]
imply
\begin{equation*}
\begin{array}{rcl}
\sin (\epsilon \, a) \, \alpha_{\1} + j \, \sinh(\epsilon \, a) \,
\gamma_{\1}   & = & \exp[ -  \nu_{\m}
 a] ~\beta_{\2} + z  \, \exp[ - \nu_{\p} a]
~\delta_{\2} + \\ & & j\, \left[ \, w \, \exp[ -  \nu_{\m}
 a] ~\beta_{\2} + \exp[ -  \nu_{\p}
 a]  ~\delta_{\2} \,
 \right]~, \\& & \\
- \epsilon \, \left[ \, \cos (\epsilon \, a) \, \alpha_{\1} + j \,
\cosh(\epsilon \, a) \, \gamma_{\1}  \, \right]  & = & \nu_{\m}
\exp[ - \nu_{\m}
 a] ~\beta_{\2} + z  \, \nu_{\p} \, \exp[ - \nu_{\p} a]
~\delta_{\2} + \\ & & j\, \left[ \, w \, \nu_{\m} \, \exp[ -
\nu_{\m}
 a] ~\beta_{\2} + \nu_{\p} \, \exp[ -  \nu_{\p}
 a]  ~\delta_{\2} \,
 \right]~.
\end{array}
\end{equation*}
Separating  the complex from the pure quaternionic part in the
previous equations and eliminating the coefficients $\alpha_{\1}$
and $\gamma_{\1}$, we obtain the following matrix equation
\begin{equation}
\left( \begin{array}{rr} \nu_{\m} \, \tan (\epsilon \, a) +
\epsilon~ & ~~~z \, \left[ \, \nu_{\p} \, \tan(\epsilon \, a) +
\epsilon \, \right]
\\
w \, \left[ \, \nu_{\m} \, \tanh(\epsilon \, a) + \epsilon \,
\right] & \nu_{\p} \, \tanh(\epsilon \,a) + \epsilon~
\end{array} \right) \, \left( \begin{array}{c}\exp[- \nu_{\m}\,
a] ~\beta_{\2} \\  \exp[ - \nu_{\p} \, a] ~\delta_{\2}
\end{array} \right)=0~.
\end{equation}
This matrix equation has a non-trivial solution only if the
determinant of
\[
\left( \begin{array}{rr} \nu_{\m} \, \tan (\epsilon \, a) +
\epsilon~ & ~~~z \, \left[ \, \nu_{\p} \, \tan(\epsilon \, a) +
\epsilon \, \right]
\\
w \, \left[ \, \nu_{\m} \, \tanh(\epsilon \, a) + \epsilon \,
\right] & \nu_{\p} \, \tanh(\epsilon \,a) + \epsilon~
\end{array} \right)
\]
vanishes, i.e.
\begin{equation}
\label{detzw} z\,w=\frac{\left[ \, \nu_{\m} \, \tan (\epsilon \,
a) + \epsilon \, \right] \, \left[ \, \nu_{\p} \, \tanh(\epsilon
\,a) + \epsilon \, \right]}{\left[ \, \nu_{\m}\, \tanh(\epsilon \,
a) + \epsilon \, \right] \, \left[ \, \nu_{\p} \, \tan(\epsilon \,
a) + \epsilon \, \right]}~.
\end{equation}
This equation can be rewritten in a more convenient form, i.e.
\begin{eqnarray}
\label{det2} \tan (\epsilon \, a)  & = &- \,
\frac{\epsilon}{\nu_{\m}} \, \, \frac{ \left( \, \nu_{\p} - z\,w
\, \nu_{\m} \, \right) \, \tanh (\epsilon \,a ) + \left( \, 1 -
z\,w \, \right) \, \epsilon }{\nu_{\p} \, \left( \, 1 - z\,w \,
\right) \, \tanh (\epsilon \,a ) + \left( \, 1 - z\,w \,
\nu_{\p}/\nu_{\m}\, \right) \, \epsilon }\,  \nonumber \\ \nonumber\\
 & = & - \, \mbox{\small $\displaystyle{\frac{\epsilon}{
\sqrt{\nu_c^{\2} - \sqrt{\epsilon^{\4} - \nu_q^{\4}}}}}$}\,\times \nonumber\\
 & &\mbox{\small $\displaystyle{
\frac{ \left[ \sqrt{\nu_c^{\2} + \sqrt{\epsilon^{\4}
-\nu_q^{\4}}}- \displaystyle{\frac{\nu_q^{\4}}{\left(\epsilon^{\2}
+\sqrt{\epsilon^{\4} - \nu_q^{\4}}\right)^{\2}}} \,
\sqrt{\nu_c^{\2} - \sqrt{\epsilon^{\4} - \nu_q^{\4}}} \right] \,
\tanh (\epsilon \,a ) + \left[ 1-\displaystyle{
\frac{\nu_q^{\4}}{\left(\epsilon^{\2} +\sqrt{\epsilon^{\4} -
\nu_q^{\4}}\right)^{\2}}} \right] \,\epsilon}{ \sqrt{\nu_c^{\2} +
\sqrt{\epsilon^{\4} -\nu_q^{\4}}}\, \left[1 -
\displaystyle{\frac{\nu_q^{\4}}{\left(\epsilon^{\2}
+\sqrt{\epsilon^{\4} - \nu_q^{\4}}\right)^{\2}}} \, \right] \,
\tanh (\epsilon \,a ) + \left[ 1-\displaystyle{
\frac{\nu_q^{\4}}{\left(\epsilon^{\2} +\sqrt{\epsilon^{\4} -
\nu_q^{\4}}\right)^{\2}}\, \sqrt{\frac{\nu_c^{\2} +
\sqrt{\epsilon^{\4} -\nu_q^{\4}}}{\nu_c^{\2} - \sqrt{\epsilon^{\4}
-\nu_q^{\4}}}}} \right] \,\epsilon}} $}\nonumber \\ \nonumber\\
 & = & f(\,\epsilon\,;\, \nu_c\,, \, \nu_q\,)\, \, ,
\end{eqnarray}
where
\[ \nu_c=\sqrt{2m\,V_{\1}/\hbar^{\2}}\hspace*{1cm}\mbox{and}\hspace*{1cm}
\nu_q=\sqrt{2m\,\sqrt{V_{\2}^{\2}+V_{\3}^{\2}}/\hbar^{\2}}\,\, .
 \]
Let us first examine the case:
$\sqrt{V_{\2}^{^{\2}}+V_{\3}^{^{\2}}}<E<\sqrt{V_{\1}^{\2}
+V_{\2}^{^{\2}}+V_{\3}^{^{\2}}}$. Observing that this condition on
the energy eigenvalues implies $\nu_q<\epsilon$ and
$\nu_c>(\epsilon^{\4} -\nu_q^{\4})^{\1 \ov \4}$, it can be
immediately seen that Eq.(\ref{det2})  represents a {\em real\,}
equation which generalizes (to the "quaternionic" case) the
well-known equation obtained in "complex" quantum mechanics, i.e.
\begin{eqnarray}
\label{detc} \tan (\epsilon \, a)  & = &f(\,\epsilon\,;\, \nu_c\,,
\,
0)\nonumber\\
 & = & - \, \frac{\epsilon}{
\sqrt{\nu_c^{\2} - \epsilon^{\2}}}\, \, .
\end{eqnarray}
One may ask wheter this is still true for energy values below
$\sqrt{V_{\2}^{^{\2}}+V_{\3}^{^{\2}}}$. For
$E<\sqrt{V_{\2}^{^{\2}}+V_{\3}^{^{\2}}}$, it can be immediately
seen that $\nu_{\pm}$ and $zw$ are complex and this seems to be
too restrictive. Consequently, no bound states should be expected
for $E<\sqrt{V_{\2}^{^{\2}}+V_{\3}^{^{\2}}}$. Nevertheless, by
simple algebraic manipulations (see appendix B for the explicit
calculations), it can be shown that Eq.(\ref{det2}) {\em still}
represents a {\em real} equation and bound states could exist
below the pure quaternionic potential. Graphical solutions of
equation (\ref{det2})  give the  energies of the bound states of a
particle in a quaternionic spherically symmetric potential trap,
see Fig.1 and Fig.2. It is also instructive to compare the
quaternionic bound states with the bound states of the
trial-complex potential
$\sqrt{V_{\1}^{^{\2}}+V_{\2}^{^{\2}}+V_{\3}^{^{\2}}}$.

\section*{IV CONCLUSIONS}
This work was intended as an attempt at motivating the study of
quaternionic formulations of quantum mechanics and at looking for
deviations from the standard complex theory. The analysis
presented in this paper represents a {\em preliminary} discussion
on quaternionic bound states and touched only a few aspects of the
theory. In particular, no attempt has been made here to develop a
quaternionic perturbation theory or to study the behavior of
confined quaternionic wave packets. These topics exceed the scope
of this paper but surely represent important points to be
investigated in view of a complete understanding of how and where
quaternionic potentials could be seen and {\em if\,} they really
exist in nature.

\newpage

\section*{Appendix A }
To calculate the radial solution in region II, we have to solve a
second order differential equation with quaternionic constant
coefficients,
\begin{equation}
\label{radialU2} \left\{ \, i \, \frac{\hbar^{\2}}{2m} \,
\frac{\mbox{d}^{\2}}{\mbox{d}r^{\2}} - i \, V_{\1} - j \, V_{\2} -
k \, V_{\3} \, \right\} \, U_{\II,\H}(r) = - \, U_{\II,\H}(r) \, i
\, E~.
\end{equation}
We refer the reader to \cite{QDO,QDO2} for a detailed analysis of
quaternionic differential operators. In this appendix, we will
touch only a few aspect of the theory, by restricting our
attention to differential operators with quaternionic constant
coefficients. A (right-complex linear) solution of
Eq.(\ref{radialU2}) can be written in terms of (left-acting)
quaternionic ($q$) and complex ($\nu$) coefficients, i.e.
\[ q \, \exp[\nu \,r]\, \, .
\]
To determine this coefficients, let us apply to
Eq.(\ref{radialU2}) the anti-Hermitian  operator
\[
 i \, \frac{\hbar^{\2}}{2m} \,
\frac{\mbox{d}^{\2}}{\mbox{d}r^{\2}} - i \, V_{\1} - j\, V_{\2} -
k \, V_{\3}\, \, .
 \]
In this way  we obtain
\begin{equation}
\label{2times}
 \left\{ - \left( \frac{\hbar^{\2}}{2m} \right)^{\2} \,
\frac{\mbox{d}^{\4}}{\mbox{d}r^{\4}} \, +\, 2 \,
\frac{\hbar^{\2}}{2m} \,  V_{\1} \,
\frac{\mbox{d}^{\2}}{\mbox{d}r^{\2}}  \, -\,  \left(
V^{^{\2}}_{\1} + V^{^{\2}}_{\2} + V^{^{\2}}_{\3} \right) \right\}
\,
 U_{\II,\H}(r) = -  \,E^{^{\2}} \,U_{\II,\H}(r)~,
\end{equation}
which represents a {\em real} differential equation. Consequently,
the quaternionic factor $q$ can be factorized and the complex
coefficient $\nu$ is calculated once solved the following
algebraic equation
\begin{equation}
 \left( \frac{\hbar^{\2}}{2m} \right)^{\2} \, \nu^{\4}  -2\,
 \frac{\hbar^{\2}}{2m} \,  V_{\1} \, \nu^{\2}  +
 V^{^{\2}}_{\1} + V^{^{\2}}_{\2} + V^{^{\2}}_{\3} -
E^{^{\2}}  = 0~.
\end{equation}
The solutions for the complex coefficients $\nu$ are given by
\begin{equation}
\label{nu} \nu_{\1,\2}=\pm \,\nu_{\m}\, \, \, \,\, \mbox{and} \,
\,\,\, \, \nu_{\3,\4}=\pm \,\nu_{\p}\, \, , \, \, \, \, \, \, \,
 \mbox{where}\, \, \,
 \mbox{$\nu_{\mip} = \sqrt{\frac{2m}{\hbar^{\2}} \, \left( V_{\1}
\pm \sqrt{E^{^{\2}} - V_{\2}{^{^{\2}}}- V_{\3}^{^{\2}} } \,
\right) }$}\, \, .
\end{equation}
Coming back to Eq.(\ref{radialU2}), for $E\neq \sqrt{
V_{\2}{^{^{\2}}}+ V_{\3}^{^{\2}} }$ and $E\neq
\sqrt{V_{\1}{^{^{\2}}}+V_{\2}{^{^{\2}}}+ V_{\3}^{^{\2}} }$ we then
find {\em four} (right-complex) linear independent
solutions\cite{QDO}, i.e.
\[
q_{\m} \, \exp[- \nu_{\m} \,r]\, \, ,  \, \, \, \, \, q_{\m} \,
\exp[ \nu_{\m} \,r]\, \, , \,  \, \,  \, \, q_{\p} \, \exp[-
\nu_{\p} \,r]\, \, \, \, \, \mbox{and} \, \, \,  \, \, q_{\p} \,
\exp[\nu_{\p} \,r]\, \, \, .
\]
To find the quaternionic factor, $q_{\m}$ (associated to $\exp[\pm
\, \nu_{\m} r]$) we set $q_{\m} = 1 + j\, w$ ($w\in \mathbb{C}$).
This choice is possible due to the right-complex linearity of the
quaternionic differential equation (\ref{radialU2}). It follows
immediately that
\[
i \, (1 + j \,w) \, \left( V_{\1} - \sqrt{E^{^{\2}} -
V_{\2}^{^{\2}}- V_{\3}^{^{\2}} } \,\right)   - \left( i \,V_{\1} +
j \,V_{\2} + k \,V_{\3} \right) (1 + j \,w) = -  \, (1 + j\, w)\,
i \, E~.
\]
It is easy to check that the previous equation implies
\begin{eqnarray}
 i \left(E - \sqrt{E^{^{\2}} - V_{\2}^{^{\2}} - V_{\3}^{^{\2}}}
\,\right) +
(V_{\2} + i \, V_{\3}) \, w = 0\, \, , \nonumber \\
i \left(E + \sqrt{E^{^{\2}} - V_{\2}^{^{\2}} - V_{\3}^{^{\2}}}\,
\right) \, w - (V_{\2} - i\, V_{\3}) = 0\, \, .
\end{eqnarray}
Finally,
\begin{equation}
w =  - \, i \, \frac{E \, - \, \sqrt{E^{^{\2}} - V_{\2}^{^{\2}} -
V_{\3}^{^{\2}}}}{ V_{\2} \, + \, i \, V_{\3}} = - \, i \,
\frac{V_{\2} \, - \, i \, V_{\3}}{E \, + \, \sqrt{E^{^{\2}} -
V_{\2}^{^{\2}} - V_{\3}^{^{\2}}}} ~.
\end{equation}
To find the quaternionic factor, $q_{\p}$ (associated to $\exp[\pm
\, \nu_{\p} r]$) we choose $q_{\p} = z + j$ ($z\in \mathbb{C}$). A
calculation similar to previous one gives
\begin{equation}
z = \, i \, \frac{E \, - \, \sqrt{E^{^{\2}} - V_{\2}^{^{\2}} -
V_{\3}^{^{\2}}}}{ V_{\2} \, - \, i \, V_{\3}} = \, i \,
\frac{V_{\2} \, + \, i \, V_{\3}}{E \, + \, \sqrt{E^{^{\2}} -
V_{\2}^{^{\2}} - V_{\3}^{^{\2}}} } ~.
\end{equation}

\newpage

\section*{Appendix B}
In this appendix, we are interested to prove that Eq.(\ref{det2})
for $E<\sqrt{V_{\2}^{^{\2}}+V_{\3}^{^{\2}}}$ {\em still}
represents a {\em real} equation. Before to proceed with the
proof, let us observe that
\[
z\,
w=\exp\left[-2\,i\,\arctan\frac{\sqrt{V_{\2}^{^{\2}}+V_{\3}^{^{\2}}-
E^{^{\2}}}}{E}\right]
\]
and
\[
\nu_{\mip} = \mbox{$\sqrt{\frac{2m}{\hbar^{\2}}} \left[
\frac{1}{\sqrt{2}} \, \sqrt{V_{\1} + \sqrt{V_{\1}^{^{\2}} +
V_{\2}^{^{\2}}+ V_{\3}^{^{\2}}- E^{^{\2}}}} \pm i
\frac{1}{\sqrt{2}} \, \sqrt{ \sqrt{V_{\1}^{^{\2}} +
V_{\2}^{^{\2}}+ V_{\3}^{^{\2}}- E^{^{\2}}} - V_{\1}} \right]$}\,
\, .
\]
It is natural to rewrite Eq.(\ref{det2}) as follows
\begin{eqnarray} \label{det2bis}
\tan (\epsilon \, a) & = &- \, \epsilon \, \, \frac{ \left( \,
\nu_{\p} - z\,w \, \nu_{\m} \, \right) \, \tanh (\epsilon \,a ) +
\left( \, 1 - z\,w \, \right) \, \epsilon }{\nu_{\m}\, \nu_{\p} \,
\left( \, 1 - z\,w \, \right) \, \tanh (\epsilon \,a ) + \left( \,
\nu_{\m}  - z\,w \, \nu_{\p}\, \right) \, \epsilon }\, \nonumber
\\
 & = & - \, \epsilon \, \, \frac{\mbox{\sc Num}}{\mbox{\sc Den}} =
 - \, \epsilon \, \, \frac{\mbox{\sc Num}\, \,\overline{\mbox{\sc Den}} }{
 \left|\mbox{\sc Den}\right|^{\2}}\, \, .
\end{eqnarray}
The important point to note here is that
\[ \nu_{\p} = \overline{\nu_{\m}}\, \,\, \, \, \, \, \, \mbox{and}
\, \, \, \, \, \, \,\, |z\,w|=1\, \, .
\]
We can now proceed with the proof,
\begin{eqnarray}
\mbox{\sc Num}\, \,\overline{\mbox{\sc Den}} & = &\left[\, \left(
\, \nu_{\p} - z\,w \, \nu_{\m} \, \right) \, \tanh (\epsilon \,a )
+ \left( \, 1 - z\,w \, \right) \, \epsilon\,\right]
\left[\,|\nu_{\m}|^{\2}\, \left( \, 1 - \overline{z\,w} \, \right)
\, \tanh (\epsilon \,a ) + \left( \, \nu_{\p}  - \overline{z\,w}
\, \nu_{\m}\, \right) \, \epsilon\,\right]\nonumber \\
 & = & \left(\,  \nu_{\m} - z\,w\, \nu_{\m} + \overline{\nu_{\m}} -
 \overline{z\,w}\,
 \overline{\nu_{\m}}\, \right)\,|\nu_{\m}|^{\2}\, \tanh^{\2} (\epsilon \,a
  )\, + \nonumber \\
   &   & \left[\, \nu_{\m}^{\,\2} + \overline{\nu_{\m}}^{\,\2} - ( z\,w +
   \overline{z\,w})\, |\nu_{\m}|^{\2} + ( 1 - z \, w)\, (1 -
   \overline{z\,w})\, |\nu_{\m}|^{\2}\,\right]\, \epsilon\,
   \tanh(\epsilon \,a
  )\, +   \nonumber \\
 &   & \left(\,  \nu_{\m} - \overline{z\,w}\, \nu_{\m} + \overline{\nu_{\m}}- z\,w\,
 \overline{\nu_{\m}}\, \right)\, \epsilon^{\2} \nonumber\\
 & \in & \mathbb{R}\,\,.
\end{eqnarray}

\newpage

\begin{figure}[hbp]
\hspace*{-2.5cm}
\includegraphics[width=20cm, height=20cm, angle=0]{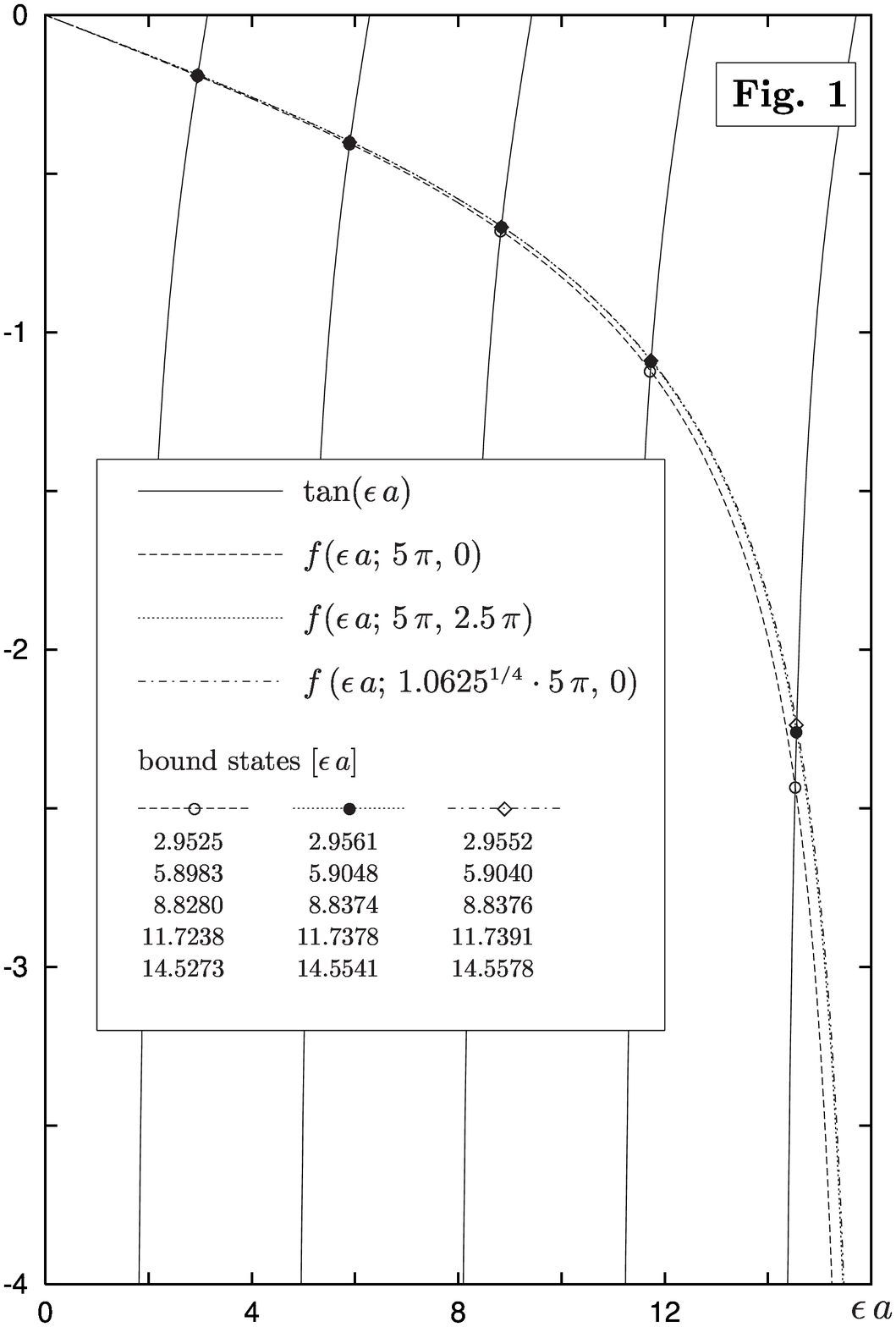}
\caption{Graphical solutions of equation (\ref{det2}). The circle,
bullet and diamond symbols respectively represent the vanishing
angular momentum bound state solutions of the radial Schr\"odinger
equation for a  complex $(\,i\,V_{\1}\,)$, quaternionic
$(\,i\,V_{\1} + j\,\,V_{\2} + k \,V_{\3}\,)$,  and trial-complex
$\left(\,i\,
\sqrt{V_{\1}^{^{\2}}+V_{\2}^{^{\2}}+V_{\3}^{^{\2}}}\,\right)$
square-well potential. For the convenience of the reader we also
list the numerical values of the bound energies. The plots refer
to the following potentials: $a \sqrt{2m\,V_{\1}/\hbar^{\2}} = 5
\, \pi$ and $a
\sqrt{2m\,\sqrt{V_{\2}^{\2}+V_{\3}^{\2}}/\hbar^{\2}} = 2.5 \,
\pi$.}
\end{figure}

\newpage

\begin{figure}[hbp]
\hspace*{-2.5cm}
\includegraphics[width=20cm, height=20cm, angle=0]{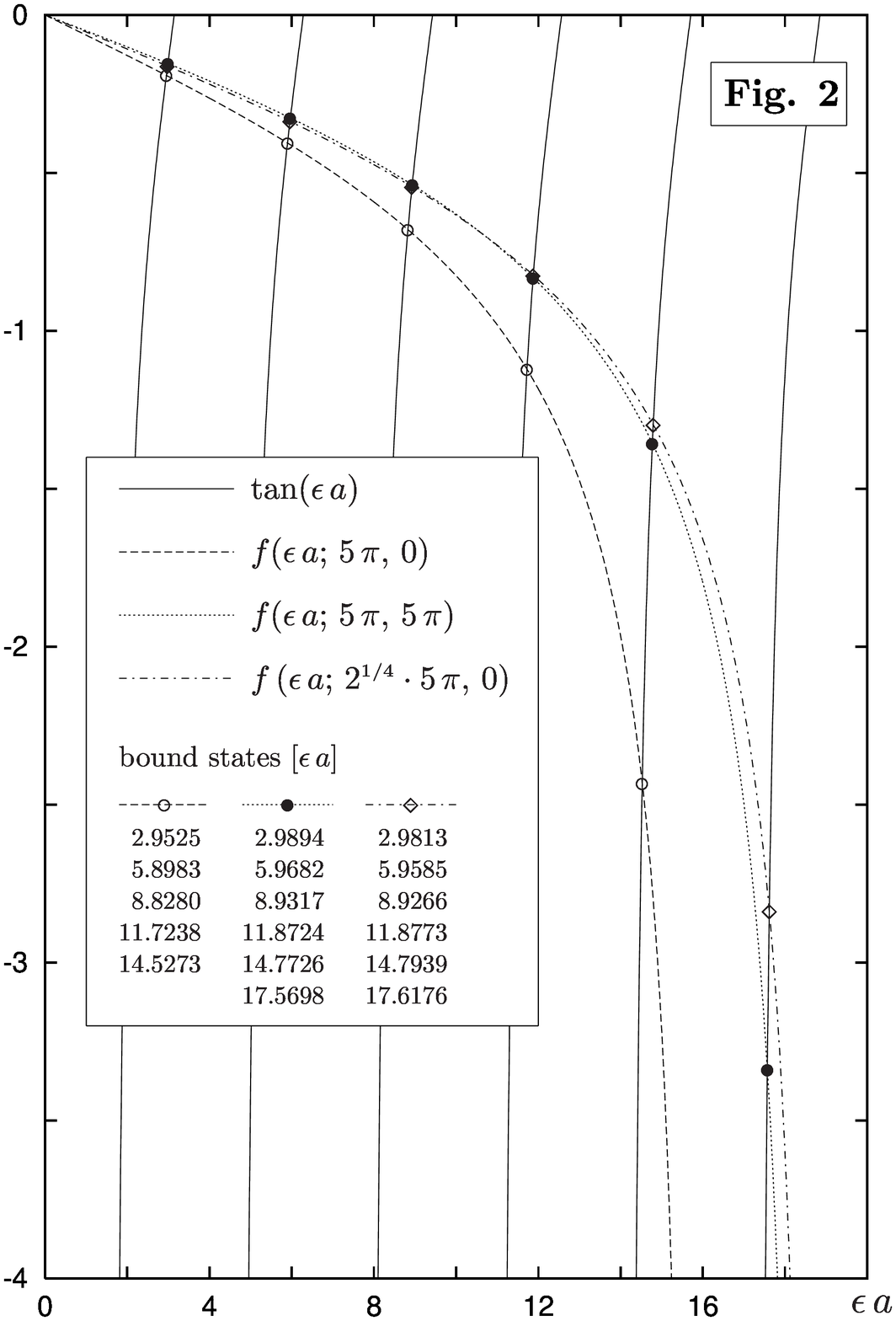}
\caption{Graphical solutions of equation (\ref{det2}). The circle,
bullet and diamond symbols respectively represent the vanishing
angular momentum bound state solutions of the radial Schr\"odinger
equation for a  complex $(\,i\,V_{\1}\,)$, quaternionic
$(\,i\,V_{\1} + j\,\,V_{\2} + k \,V_{\3}\,)$,  and trial-complex
$\left(\,i\,
\sqrt{V_{\1}^{^{\2}}+V_{\2}^{^{\2}}+V_{\3}^{^{\2}}}\,\right)$
square-well potential. For the convenience of the reader we also
list the numerical values of the bound energies. The plots refer
to the following potentials: $a \sqrt{2m\,V_{\1}/\hbar^{\2}} = 5
\, \pi$ and $a
\sqrt{2m\,\sqrt{V_{\2}^{\2}+V_{\3}^{\2}}/\hbar^{\2}} = 5 \, \pi$.
}
\end{figure}


\begin{thebibliography}{99}




\bibitem{ADL}
S. L. Adler, {\it Quaternionic quantum mechanics and quantum
fields}, (New York: Oxford University Press, 1995).

\bibitem{EP1}
S. De Leo and G. Scolarici, ``Right eigenvalue equation in
quaternionic quantum mechanics'', J. Phys. A {\bf 33}, 2971--2995
(2000).

\bibitem{EP2}
S. De Leo, G. Scolarici and L. Solombrino, ``Quaternionic
eigenvalue problem'', J. Math. Phys. {\bf 43}, 5815--5829 (2002).


\bibitem{QDO}
S. De Leo and G. Ducati, ``Quaternionic differential operators'',
J. Phys. Math.  {\bf 42}, 2236--2265 (2001).


\bibitem{DAV89}
A. J. Davies and B. H. McKellar, ``Non-relativistic quaternionic
quantum mechanics'', Phys. Rev. A {\bf 40},  4209--4214 (1989).

\bibitem{DAV92} A. J. Davies and B. H. McKellar, ``Observability
of quaternionic quantum mechanics'', Phys. Rev. A {\bf 46},
3671--3675 (1992).


\bibitem{QTE} S. De Leo, G. Ducati and C. Nishi, ``Quaternionic
potential in non-relativistic quantum mechanics'', J. Phys. A {\bf
35}, 5411--5426 (2002).

\bibitem{PER}
A. Peres, ``Proposed test for complex versus quaternion quantum
theory'', Phys. Rev. Lett. {\bf 42}, 683--686 (1979).

\bibitem{KAI}
H. Kaiser, E. A. George and S. A. Werner, ``Neutron
interferometric search for quaternions in quantum mechanics'',
Phys. Rev. A {\bf 29}, 2276--2279 (1984).

\bibitem{KLE}
A. G. Klein, ``Schr\"odinger inviolate: neutron optical searches
for violations of quantum mechanics'', Physica B {\bf 151}, 44--49
(1988).


\bibitem{CQQM}
S. De Leo and W. A. Rodrigues, ``Quantum mechanics: from complex
to complexified quaternions'', Int. J. Theor. Phys. {\bf 36},
1165-1177 (1997).


\bibitem{OQM}
S. De Leo and K. Abdel-Khalek, ``Towards an octonionic world'',
Int. J. Theor. Phys. {\bf 37}, 1945-1985 (1997).


\bibitem{CSP}
P. Rotelli, ``The Dirac equation on the quaternionic field'', Mod.
Phys. Lett. A {\bf 4}, 993--940 (1989).



\bibitem{RE1}
A. W. Conway, ``Quaternion treatment of the relativistic wave
equation'', Proc. Roy. Soc. A {\bf 162}, 145-154 (1937).



\bibitem{RE2}
D. Hestenes, ``Real spinors field'', J. Math. Phys. {\bf 8},
798-808 (1967); `Spin and isospin'', {\em ibidem}, 809-812 (1967).

\bibitem{RE3}
A. Gsponer and J. P. Hurni, ``Comment on formulating and
generalizing Dirac's, Proca's, and Maxwell's equations with
biquaternions or Clifford numbers'', Found. Phys. Lett. {\bf 14},
77-85 (2001).


\bibitem{RE4}
S. De Leo, ``Quaternionic Lorentz group and Dirac equation'',
Found. Phys. Lett. {\bf 14}, 37-50 (2001).


\bibitem{GT1}
S. De Leo and P. Rotelli, ``Quaternionic electroweak theory'', J.
Phys. G {\bf 22}, 1137-1150 (1996).



\bibitem{GT2}
S. De Leo and G. Ducati, ``Quanternionic group in physcis'', Int.
J. Theor. Phys. {\bf 38}, 2197-2220 (1999).



\bibitem{QDO2}
S. De Leo and G. Ducati, ``Solving simple quaternionic
differential equations'', J. Phys. Math.  {\bf 44}, 2224--2233
(2003).

\end{thebibliography}
\end{document}